\documentstyle[preprint,aps]{revtex}

\newcommand{\beq}{\begin{equation}}
\newcommand{\eeq}{\end{equation}}
\newcommand{\beqarr}{\begin{eqnarray}}
\newcommand{\eeqarr}{\end{eqnarray}}

\begin{document}

\title{Duality near Quantum Hall Transitions}
\author{E. Shimshoni$^{1}$, S. L. Sondhi$^{2}$ and D. Shahar$^3$}
\address{$^1$Department of Mathematics-Physics, Oranim--Haifa
University, Tivon 36006,
Israel.}

\address{$^2$Department of Physics, Princeton University, Princeton
NJ 08544}

\address{$^3$Department of Electrical Engineering, Princeton University,
Princeton NJ 08544}

\maketitle
\begin{abstract}
A recent experiment by Shahar et al \cite{expts}, on the phase transitions
between quantum Hall states and the insulator, found that the
current-voltage characteristics in the two phases are related by
symmetry. It was suggested in this work that this is evidence for
charge-flux duality near quantum Hall transitions. Here we provide
details of this analysis.
We review some theoretical ideas on charge-flux duality in the composite
boson description of the quantum Hall effect, and interpret the data
as implying that this duality is a symmetry in the transition region {\em and}
that the Hall response of the bosons vanishes.
We observe that duality for composite bosons is equivalent to a particle-hole
symmetry for composite fermions and show that a Landauer analysis of transport
for the latter allows a possible understanding of the reflection symmetry
and Hall response beyond the linear regime. We note that the duality
interpretation
supports the scenario of superuniversality for quantum Hall transitions
outlined by Kivelson, Lee and Zhang \cite{klz}.
Finally, we discuss how to search for duality at other transitions
in the quantum Hall regime.

\end{abstract}
\pacs{73.40H, 71.30}

\section{Introduction}

Two dimensional electron systems in the quantum Hall (QH) regime,
exhibit an extremely complex phase diagram with a rich set of
transitions between various phases. In recent years, a combination
of theoretical and experimental work has led to considerable
progress in understanding this complexity. An important milestone
was the work of Kivelson, Lee and Zhang (KLZ) \cite{klz} who proposed
a topology for the phase diagram (in the disorder-magnetic field
plane) on the basis of a set of correspondence rules. In the main,
these codified two ideas which are central to the current
understanding of the QHE. The first of these is the existence of sets
of QH states that are related by flux attachment transformations
that are the basis of Chern-Simons theories of the QHE \cite{chernsimons}
and of the work of Jain \cite{jain}. The second idea is the hierarchical
structure of the set of QH states, i.e. that, starting with
a basis of primary (Laughlin) states, one can construct members of the set
by ``condensing'' quasiparticles of other members of the set
\cite{hierarchy}. In the
bosonic Chern-Simons framework used by KLZ, quasiparticles are vortices
and hence the hierarchical descendents are obtained as saddle points of
actions obtained by repeated duality transformations. The topology of
the phase diagram is then fixed by requiring that neighboring
phases be related by the condensation of {\em one} set of quasiparticles,
or by one duality transformation. Experimentally, this expectation
has been borne out fairly well and although there is evidence of systematic
departures at low fields and in the reentrant region near $\nu=1/5$
\cite{exceptions}, these are not at issue in the region of interest to us
in this paper.

The transitions between the various phases in the QH regime have been
the focus of another active line of work, particularly following
the contributions of Pruisken \cite{pruisken} and Wei, Tsui, Paalanen
and Pruisken \cite{wei1} who proposed that the transitions between
integer QH states are quantum phase transitions, i.e. continuous
$T=0$ phase transitions. Subsequent experimental work has supported this
identification for transitions between other QH states and has suggested
a hypothesis of ``super-universality'', i.e. that {\em all} such transitions
share the same exponents, and a tentative identification of the correlation
length exponent $\nu_\xi$ and the dynamic scaling exponent $z$
\cite{fracins,rmp}; remarkably, the value of $\nu_\xi$ agrees with
numerical calculations for non-interacting electrons in the lowest Landau
level \cite{bodo}. In the work of KLZ, an explanation was offered for the
superuniversality
(along with an extension to the QH/insulator transition of interest
to us) in terms of a gaussian analysis of the fluctuations of the Chern-Simons
field which allowed them to predict a set of critical conductivities for
the various transitions as well. While the latter predictions have gathered
some experimental support, their neglect of higher order fluctuations
has been criticized and defended in model calculations \cite{against,for}.
In Section IV we comment on the issues in this debate.

A recent experiment of Shahar {\em et al.}\cite{expts}, on the transition
between the $1/3$ QH fluid and the proximate, high field insulator,
appears to tie these
two streams of ideas together in a very interesting manner while also
lending strong support to
KLZ's account of the critical behavior. To recapitulate, the experiment
found that for a range of fields near the transition between the
QH state and the insulator,
the transport data in the two phases are related by an unanticipated
symmetry. More precisely, there exist filling fractions
$(\nu, \nu_d)$ in the QH and insulating states at which the {\em non-linear}
longitudinal current density/electric field ($j, E_{L}$) characteristics
are related by {\em reflection}:
\begin{equation} \label{eq:long}
\{j(\nu_d), E_{L}(\nu_d)\} =
\{ {e^2 \over h} E_{L}(\nu), {h \over e^2} j(\nu) \} \ .
\end{equation}
On the basis of the functional relationship between $\nu_d$ and $\nu$,
this was identified in \cite{expts} as a manifestation of charge-flux
duality for composite bosons and equivalently, particle-hole symmetry
for composite fermions. This was then used to argue that the Hall response
must also be symmetric and have the particularly simple form of the
($j, E_{H}$) characteristics being {\em linear} and unchanged across the
transition,
\begin{equation}\label{eq:hall}
E_H = 3 {h \over e^2} j \ \ \ \forall \nu \ ,
\end{equation}
an expectation borne out by the data presented in \cite{expts}.
Similar results hold at the $\nu=1$ to insulator transition \cite{expts2},
implying that the symmetry is present at the transitions between
primary QH fluids and the insulator. (In the rest of the paper, we
shall use {\em reflection symmetry} as a shorthand for the experimentally
observed symmetry of the response summarized in Eqs. (\ref{eq:long}) and
(\ref{eq:hall}) while reserving the term duality for the theoretical
inference about the underlying dynamics.)

In this paper we present the theoretical underpinnings of the interpretation
advanced in \cite{expts}. In Sec. 2 we
describe these in language of composite
bosons with the help of the Chern-Simons resistivity addition law and duality
transformations. We comment there on the issue of whether the critical action
is gaussian. Next, in Sec. 3 we introduce a Landauer formulation of the
transport in the
composite fermion basis that connects the experimental observations to
particle-hole symmetry for the fermions---which is shown to be equivalent
to duality for the bosons. In both cases we note that an additional postulate,
equivalent to the constancy of the Hall response, needs to be introduced
separately.

In Sec. 4 we switch gears,
and adduce purely theoretical evidence that this symmetry
is present at long wavelengths by revisiting existing studies of QH
transitions. In Sec. 5, we discuss how the symmetry is hidden at
inter-plateaux transitions and show how it can be detected by a deconvolution
of the experimental data. We end with a summary of our argument and some
key open issues in our approach. As our work draws on much previous work
in addition to that reviewed already, most notably that Fisher
\cite{mpaf-dual} and of Ruzin and collaborators \cite{Ruzin1,Ruzin2},
we comment on these connections in the main text.

\section{The Bosonic Approach -- Duality}
\label{sec:boson}
\subsection{Dual Filling Factors}

We begin by recapitulating the composite boson interpretation of the transition
between the primary Hall states and the insulator. A Chern-Simons
transformation \cite{klz,REF:ZHK} is used to represent the physical electrons
as bosons
that carry $k$ flux quanta ($\phi_0$), $k$ being an odd integer. At the filling
factor $\nu$, the resulting bosons experience a mean magnetic field
$B_{eff}=B(1-k \nu)$ which vanishes at $\nu=1/k$ thus enabling them to condense
into a superconducting state. A decrease in $\nu$ results in a non-zero
$B_{eff}$ which induces vortices. At first the vortices are trapped by disorder
and the bosons continue to superconduct. However at a critical density the
vortices delocalize and destroy the superconductivity of the bosons. This
causes
the bosons to become localized and the original electronic system becomes
insulating as well. The composite boson picture thus naturally associates the
physics of QH/Insulator transitions with that of field-tuned
superconductor/insulator
transitions \cite{SITrev}.

This description is clearly marked by a {\em duality} in the roles of the
bosons/vortices which are localized in the insulating/superconducting phases
respectively and are delocalized in the conjugate phase. This suggests that
the problem might be characterized by a {\em symmetry}, which we shall also
refer to as duality \cite{fn-dual}, that relates the superconducting and
insulating
phases in that the dynamics of the system expressed in terms of the bosons
at a filling $\nu$ in the QH phase, is identical to the dynamics at $\nu_d$
in the insulating phase expressed in terms of the vortices. The natural
candidates for the dual fillings are related by,
\begin{equation}
\frac{n_v(\nu)}{n_b(\nu)} = \frac{n_b(\nu_d)}{n_v(\nu_d)},
\label{bosenu}
\end{equation}
where $n_b$ and $n_v$ are the boson and vortex densities respectively. As
the vortex density is set by the effective field, $\rho_v = B_{eff}/\phi_0
= \rho_b ( \nu^{-1} - k)$, it follows that
\begin{equation}
\frac{1}{\nu^{-1} - k} = \nu_d^{-1} - k \ .
\label{bosnunud}
\end{equation}
Explicitly, $\nu_d = (1-k\nu)/[\nu + k(1-k\nu)]$ \cite{expts}. The observation,
that the filling fractions related by the reflection symmetry in the
experiment,
obey Eq. (\ref{bosnunud}) to a good approximation, strongly suggests that the
reflection symmetry is a manifestation of duality.

To further this interpretation, we need to examine whether the reflection
symmetry follows from an assumption of identity between the boson
and vortex actions at $\nu$ and $\nu_d$, respectively.  In order to do
this we rederive the well known Chern-Simons resistivity relation
\cite{ioffetc},
that relates the bosonic response to the electronic  response, in
a way that suggests its validity beyond the linear regime.

\subsection{Chern-Simons Resistivity Addition}

Our starting point is the bosonic Chern-Simons functional integral
\cite{REF:ZHK} for the ``Helmholtz'' functional $F[A_\mu;\nu]$ \cite{freeE},
of the electrons at filling fraction $0\leq\nu\leq 1/k$,
in the presence of an external 3-vector gauge field $A_\mu$ ($\hbar=e/c=1$),
\begin{eqnarray}
e^{-F[A_\mu;\nu]}&=&\int D[a_\mu] e^{-\frac{i}{2k}
S_{cs}}e^{-F^b[A_\mu+a_\mu;\nu]}
\nonumber \\
S_{cs}&\equiv& \int d{\bf r}d\tau \epsilon_{\mu\nu\lambda}
a_\mu\partial_\nu a_\lambda\; .
\label{EQ:fefb}
\end{eqnarray}
Here,
\beq
e^{-F^b[A_\mu+a_\mu;\nu]} = \int D[\phi] e^{-S_{M}(\phi, A_\mu+a_\mu; \nu)}
\eeq
defines the free energy of the composite bosons, which is
obtained after (exactly) integrating out the fluctuations of the boson
field.
The observable electro-magnetic response (associated with functional
derivatives of $F$) is thus related to the response
of the bosons to the effective field $A_\mu+a_\mu$.
In linear response,
the free energies are quadratic in the gauge fields, with coefficients
that are directly related to the conductivity tensor computed in RPA.
The integration in Eq.~(\ref{EQ:fefb}) can be performed explicitly,
resulting in an algebraic relation between $\sigma_{ij}$, $\sigma^b_{ij}$
of the physical electrons and bosons, respectively. This relation is most
simply phrased in terms of the {\it resistivity} tensors. Here we proceed
slightly differently, in a manner which might generalize to non--linear
response.

Our strategy is to formally introduce the resistivity tensor {\it before}
carrying out the integration over $a_\mu$ in Eq.~(\ref{EQ:fefb}). We convert
from the Helmholtz functional $F[{\bf A};\nu]$ to the ``Gibbs'' \cite{freeE}
functional $G[{\bf J};\nu]$ via a functional Fourier transform:
\beq
e^{-G[{\bf J};\nu]}=\int D[{\bf A}]e^{-F[{\bf A};\nu]}
e^{-i\int {\bf A}\cdot {\bf J}} \;.
\label{EQ:gibbs}
\eeq
(For quadratic actions, this can substitute for the standard Legendre
transform and therefore allows a purely functional integral argument.
It is unclear whether something of this kind can be done beyond
quadratic order---this will be further discussed elsewhere \cite{SSnlr}.)
Hereon we choose the gauge $A_0=0$ for all vector potentials, so that the
corresponding electric fields are (in Fourier space) $E_j=i\omega A_j$.
To quadratic order, the functionals $F$ and $G$ are given by
\begin{eqnarray}
F[{\bf A};\nu]&=&\int d^3 q{1\over 2}\sum_{i,j}\omega A_i(q)
\sigma_{ij}(q;\nu)A_j(-q) \nonumber \\
G[{\bf J};\nu]&=&\int d^3 q {1\over 2}\sum_{i,j}(1/\omega)
J_i(q)\rho_{ij}(q;\nu)J_j(-q)
\end{eqnarray}
where $q=(\omega,{\bf q})$ and $i,j \epsilon \{x,y\}$.
In the same way, we define $G^b[{\bf J}]$ for the bosons.
We then insert Eq.~(\ref{EQ:fefb})
into the definition of $G[{\bf J}]$, Eq.~(\ref{EQ:gibbs}).
Integrating over $A_\mu+a_\mu$, and using the corresponding definition of
$G^b[{\bf J}]$, this yields
\begin{eqnarray}
e^{-G[{\bf J};\nu]}&=&e^{-G^b[{\bf J};\nu]}
\int D[{\bf a}]e^{iS[{\bf a}]} \\ \nonumber
S[{\bf a}]&\equiv& \int d^3q\left\{\left({i\omega\over k}\right)
a_x(-q) a_y(q)+ {\bf a}(-q)\cdot {\bf J}(q)\right\}\; .
\label{EQ:gibbs1}
\end{eqnarray}
The integration over {\bf a} then yields
\beq
G[{\bf J};\nu]=G^b[{\bf J};\nu]+
\int d^3q(k/\omega)J_x(q) J_y(-q)\, .
\label{EQ:gbge}
\eeq
It follows that the DC resistivities of the electrons and bosons
are related by the addition of the Chern-Simons resistivity, i.e.
\beq
\rho = \rho^b + \rho^{\rm cs}(k) \ ,
\label{EQ:rerb}
\eeq
where
\begin{eqnarray}
\rho &=& \pmatrix{\rho_L & \rho_H \cr -\rho_H & \rho_L  \cr}
\nonumber \\
\rho^b &=& \pmatrix{\rho^b_L & \rho^b_H \cr -\rho^b_H & \rho^b_L  \cr}
\nonumber \\
\rho^{\rm cs}(k) &=& \pmatrix{0 & k  \cr -k & 0  \cr}
\end{eqnarray}
are resistivities scaled by $h/e^2$. (We follow this convention hereafter.)
This is frequently summarized by the statement that the  measured voltage
is the sum of the bosonic response and the Faraday effect stemming from the
flux carried by them. This last interpretation is clearly not
limited to linear response and hence we might expect Eq. (\ref{EQ:rerb})
and some version of Eq. (\ref{EQ:gbge}) to hold in the non-linear region
as well.
(This extension can also be argued by proceeding from Ehrenfest's theorem
applied to the operator equations of motion for the Chern-Simons Lagrangian
of \cite{REF:ZHK}. The real problem here, discussed further in
Section \ref{ph-land}, is the lack of a fluctuation-dissipation theorem
beyond the linear regime.)

In proceeding from Eq. (\ref{EQ:fefb}) to Eq. (\ref{EQ:rerb}), we have ignored
a number of delicate issues regarding disorder averages and the order of the
$k$,
$\omega$ and $T$ limits. On the latter point we have in mind the ordering $k
\ll
\omega \ll T$ so that we are in the correct regime for defining transport
coefficients
and mesoscopic fluctuations are not an issue on account of dephasing proceses
at $T \ne 0$. Consequently, our functional integrals are implicitly bounded
between $\tau=0$ and $\tau=\hbar \beta$. Giving meaning to the boson transport
coefficients is more subtle. Outside of our formal manipulations with
functional
integrals, the best approach is to think of the conductivities as being given
diagrammatically by the sum of graphs that are one-particle irreducible with
respect to the Chern-Simons interaction but are permitted to have an arbitrary
set of internal Chern-Simons, disorder and scalar interaction
lines. With this definition, Eq. (\ref{EQ:rerb}) is a tautology.

\subsection{Duality and Transport: Phenomenology}

We now return to our main argument and consider the implications of
duality for $\rho^b$; in the next section we will interpret the
reflection symmetry in their light.
Our discussion here  is a generalization
to the non-linear case of Fisher's discussion for the
field tuned superconductor insulator transition\cite{mpaf-dual}.
By hypothesis, the boson and vortex dynamics are the same at dual
fillings except that they see oppositely directed magnetic fields.
Hence the boson and vortex resistivity matrices at these fillings
are related by,
\beq
\rho^b({j},\nu) = [\rho^v({j_v},\nu_d)]^\dagger \ ,
\eeq
where the explicit dependence on the boson and vortex currents
indicates the possibly non-linear nature of the response.
As bosons and vortices see electric fields and currents
oppositely, their resistivity and conductivity matrices at
the {\em same} filling are related by
\beq
\rho^v({j_v},\nu_d) = \sigma^b({E}, \nu_d) \ .
\eeq
Combining these we find,
\beq\label{eq:dualconst}
\rho^b({j},\nu) = [\sigma^b({E}, \nu_d)]^\dagger \ ,
\eeq
which summarizes the implications of duality for the transport
coefficients in and beyond the linear regime.

In the linear regime, Eq. (\ref{eq:dualconst}) can be written
explicitly as a pair of relations between the resistivities,
\begin{eqnarray}\label{eq:dualinear}
\rho_{L}^b(\nu) &=& \frac{\rho_{L}^b(\nu_d)}{[\rho_{L}^b(\nu_d)]^2 +
[\rho_{H}^b(\nu_d)]^2} \nonumber \\
\rho_{H}^b(\nu) &=& \frac{\rho_{H}^b(\nu_d)}{[\rho_{L}^b(\nu_d)]^2 +
[\rho_{H}^b(\nu_d)]^2} \ .
\end{eqnarray}
We see that in general, these relate either $\rho^b_{L}(\nu)$ or
$\rho_{H}^b(\nu)$ to {\em both} $\rho^b_{L}(\nu_d)$ and
$\rho_{H}(\nu_d)$, i.e. duality does not always preserve the
longitudinal/Hall distinction. (At the self-dual critical point,
$\nu_d=\nu$, Eqs.~(\ref{eq:dualinear}) lead to the constraint
$(\rho_{L}^b)^2 + (\rho_{H}^b)^2 =1$ on the critical resistivities.)

\subsection{Duality and Transport: Experiments}

Turning now to the experimental data, we note that Eqs. (\ref{eq:long})
and (\ref{eq:hall}) imply that the longitudinal bosonic response
obeys
\beq
\rho^b_{L}(\nu,{j}) = \sigma^b_{L}(\nu_d,{E})
\eeq
(${j}$ and ${E}$ have the same numerical value in our units),
{\em and} that the bosonic Hall response vanishes, i.e.
\beq
\rho^b_{H}(\nu,{j}) = 0
\eeq
for all $\nu$ in the transition region.
These evidently satisfy the constraint in Eq. (\ref{eq:dualconst}) in
a particularly simple way. It follows then that it is necessary
but not sufficient to postulate duality in order to account for the
data---but duality and a vanishing bosonic Hall response are
both necessary and sufficient.

Let us briefly speculate on possible constraints on the form of the critical
point action from the reflection symmetry. If  Eq. (\ref{EQ:gbge}) could
be placed on a secure footing, the reflection symmetry would suggest that
the bosonic Gibbs functional, defined as the Legendre transform of the
bosonic Helmholtz functional, has the separable form,
\beq
G^b[{\bf J}]=G^b_l[J_x]+G^b_l[J_y] \ ,
\eeq
with the duality constraint
\beq
G^b_l[J_j;\nu]=F^b_l\left[{1\over i\omega}J_j;\nu_d\right] \ .
\eeq
At the critical point this would require a functional form invariant under
a Legendre transform, i.e. a gaussian.

\subsection{Duality and Transport: Duality Transformations}
\label{sec:dualdual}

In our discussion thus far, we have relied upon the
phenomenology of bosons and vortices familiar from the context of
superfluidity/superconductivity and the resistivity addition law
with a fixed Chern-Simons coefficient. Now we will present a
derivation of Eqs. (\ref{eq:dualinear})
from the viewpoint of a duality transformation in which we trade the
bosons and Chern-Simons field for another set of bosons and a different
Chern-Simons field. In doing so we will make a crucial assumption,
which is less restrictive than one made by KLZ in their original
analysis but is very much in their spirit, which will thus be seen to
be equivalent to the assumption of duality.

The formulation of the duality transformation that we will need
is the following. The bosonic Chern-Simons path integral expression
for the partition function of the electron gas at filling fraction $\nu$
\beq\label{eq:original}
Z[A_\mu] =\int D[a_\mu] \int D[\phi] e^{- \frac{i}{2k} S_{cs}(a_\mu)
- S_M(\phi, A_\mu+a_\mu; \nu)}
\eeq
is believed to be rewritable \cite{lee-fisher}, up to irrelevant terms,
as another bosonic path integral that differs only in the Chern-Simons
coefficient, the charge of the bosons and their filling fraction,
\beq\label{eq:dual}
Z[A_\mu] = e^{ -\frac{i}{2k} S_{cs}(A_\mu)}
\int D[\tilde{a}_\mu] \int D[\tilde{\phi}] e^{-i \frac{k}{2}
S_{cs}(\tilde{a}_\mu)- S_M(\tilde{\phi}, \frac{1}{k}A_\mu+
\tilde{a}_\mu; \nu_d)}
\ .
\eeq
Note that the transformed bosons, which are the vortices, conduct in
parallel to the ideal QH fluid with $\sigma_{xy}=1/k$.

It follows from Eq. (\ref{eq:original}) that the resistivity at $\nu$
takes the form
\beq\label{eq:add2}
\rho(\nu)=\rho^b(\nu) + \rho^{cs}(k)
\eeq
where $\rho^b$ is the boson resistivity. Similarly it follows from
Eq. (\ref{eq:dual}) that the conductivity at $\nu_d$ can be written as
\beq
\sigma(\nu_d) = \sigma^{cs}(k) +
\lbrace k^2 \tilde{\rho}^b(\nu) + k^2 \rho^{cs}(-k)\rbrace^{-1}
\eeq
where $\sigma^{cs}(k)=[\rho^{cs}(k)]^{-1}$ is the Chern-Simons conductivity
and $\tilde{\rho}^b$ is the {\em dual} boson resistivity, i.e. the
resistivity of the bosons in the dual representation. If we make
the assumption that this is the {\em same} function of
filling fraction as $\rho^b$ we find that $\rho(\nu_d)$
can be rewritten in the form of Eq. (\ref{eq:add2}) with $\rho^b(\nu_d)
=[\sigma^b(\nu)]^\dagger$ as before. We emphasize that this is not a
trivial assumption---the two sets of bosons interact with gauge fields
governed by Chern-Simons terms with different coefficients and therefore
their resistivities might be expected to exhibit different functional
dependences on $\nu$. Of course, the requirement that the $\rho(\nu)$
be the same in both representations does relate $\rho^b(\nu)$ with
$\tilde{\rho}^b(\nu_d)$.

\section{The Fermionic Approach -- Particle-Hole Symmetry}

\subsection{Symmetric Fillings and Equivalence with Duality}

We will now consider the QH/Insulator transitions in the composite
fermion description \cite{jkt}. In this, the electronic state
at $\nu$ is related by an even flux attachment (Chern-Simons)
transformation to a state of composite fermions at the auxiliary
filling fraction $\nu' = \nu/[1-(k-1)\nu]$; here $k-1$, with $k$ odd,
is the number of flux quanta. This maps $\nu=1/k$ to $\nu'=1$ and
the transition to the insulator has the form of the depletion of
a single filled (pseudo) Landau level, i.e. the $\nu=1 \rightarrow 0$
transition. (However, the transitions are {\em prima facie} different,
as the $k\ne1$ transitions involve gauge fields not present in the
$k=1$ case.)

As the space of states involved in the transition now includes a
full Landau level, it becomes possible to formulate a particle-hole symmetry
for the fermions that asserts that the dynamics of the particles at
auxilliary filling $\nu'$ is the same as that of holes at auxilliary
filling $\nu_d'=1-\nu'$.
In fact, this symmetry is the same as duality for the composite bosons as can
be seen by rewriting the particle-hole condition in terms of the filling
factor itself and noting its equivalence with the duality condition
derived earlier. Again, without addressing the microscopic origin of
the symmetry, we will explore its consequences for the electronic
transport. In the following we will argue that it implies the reflection
symmetry, provided we accept a particular Landauer framework for
the analysis of the transport.

\subsection{Particle-Hole Symmetry and Transport}
\label{ph-land}

Much as in the bosonic description, transport coefficients for the
electrons and the fermions are related by another addition relation,
\beq\label{eq:addf}
\rho=\rho^f(\nu) + \rho^{cs}(k-1)
\eeq
where $\rho^f$ is the fermion resistivity. Before turning to the
implications of particle-hole symmetry, we will  sketch a
Landauer theory for $\rho^f$ along the lines of Jain and
Kivelson's treatment of non-interacting electrons \cite{jk}.
Our motivation here is twofold. First, we take the experiments to
date, with their surprising estimate of a correlation length
exponent consistent with non-interacting calculations, as suggesting
that a composite fermion quasiparticle description continues to hold
even in the transition regions so that we can use an effective single
particle description for them.
Next, we feel that if the experiments are measuring universal
transport data in a critical region, it should not matter exactly
how we compute these quantities; i.e. the universal part of the
transport might be computable by our idealized Landauer calculation
even if there are non-universal parts sensitive to the actual
arrangements of contacts in the device. Needless to say, this
assumption needs further study \cite{lopez}.

With these caveats, we consider the conductances of a disordered QH
region sandwiched between two ideal regions that serve to define
incoming and outgoing edge states with linear dispersions
\cite{fn-edgedisp}. We imagine a calculation in the critical  region
where the size of the disordered region is set by a dephasing length.
The transport through the region is
characterized by an energy dependent transmission coefficient
${\cal T}(\epsilon)$. Following Jain and Kivelson, we define the current
$I$, longitudinal voltage $V_{L}$ and Hall voltage $V_{H}$
for given edge chemical potentials $\mu_1=\epsilon_F - V/2$ and
$\mu_2 = \epsilon_F + V/2$ ($\epsilon_F$ is the equilibrium chemical
potential on the both edges) as
\beqarr
I &=& \int_{\mu_L}^{\mu_R}d\epsilon \; {\cal T}(\epsilon) \nonumber \\
V_{L} &=& (\mu_2 - \mu_1) - \int_{\mu_L}^{\mu_R}d\epsilon \; {\cal T}(\epsilon)
\nonumber \\
V_{H} &=& \int_{\mu_1}^{\mu_2}d\epsilon \; {\cal T}(\epsilon) \ .
\label{eq:ivlvh}
\eeqarr
Evidently, these expressions allow for non-linear dependences of
the currents and voltages on the source-drain voltage $V=\mu_2 - \mu_1$
and hence on each other; all that this required is for ${\cal T}(\epsilon)$
to have structure on the scale of $V$. In the transition region, this
is easily arranged for the variation of ${\cal T}(\epsilon)$ becomes
arbitrarily rapid
at low temperatures (it is a step function at zero temperature).
Consequently, one should generically expect non-linear transport from
the proximity to the critical point.

Nevertheless, the most striking feature of Eqs. (\ref{eq:ivlvh}) is
the linear dependence of $V_H$ on $I$ independent of $\epsilon_F$ and
hence of $\nu$. This striking feature of the experimental data follows
automatically from Jain and Kivelson's Landauer formalism. If, in addition,
we postulate particle-hole symmetry ($\epsilon_c=0$),
\beq
{\cal T}(\epsilon) = 1-{\cal T}(-\epsilon)
\eeq
we find that in going between symmetry related fillings, $\epsilon_F
\rightarrow -\epsilon_F$, $V_L$ and $I$ simply trade places. Such a
treatment can then account naturally for the symmetry even in the
non-linear transport.

However, an important caveat is in order. The Landauer formalism assumes
that the dissipation necessary to produce steady state transport takes
place in the reservoirs defining
the edge chemical potentials and hence the non-linearities it produces
are due to elastic physics alone. In a real system one needs to worry
about non-linearities arising from dissipative bottlenecks, possibly
coming from critical physics themselves \cite{girvin}.
In other words, there is no fluctuation-dissipation theorem for non-linear
response which relates it uniquely to equilibrium correlations and,
in principle, it is necessary to include explicit dissipative mechanisms
in calculations.

Evidently, we have not done so in our treatment of the non-linear response
in the bosonic description and have chosen a particular, infinitely
efficient, mechanism in the fermionic Landauer description. For our purposes,
the importance of the latter is that it is a proof of principle that
intrinsic critical physics {\em can} lead to the non-linear symmetry
observed in the data.

\section{Theoretical Evidence for Duality/Particle-Hole Symmetry
and a Constant $\rho_{H}$}

Our discussion so far has focused on phenomenology in that we have
attempted to translate the experimental observations into the framework
of the Chern-Simons description of the QH transitions. This has led
us to conclude the the composite boson description must be marked
by duality and a vanishing Hall response and the composite fermion
description by particle-hole symmetry and a constant Hall response.
In this section we will review previous theoretical work which,
though it did not anticipate the particular striking features of the
data, does suggest that our inferences would arise naturally in a
microscopic theory of the QH phase transitions.

\noindent
{\bf Duality:} In Section \ref{sec:dualdual} we noted the belief, based
on the work of Lee and Fisher \cite{lee-fisher}, that the bosonic Chern-Simons
action with coefficient $k$ for particles of charge $1$ at filling fraction
$\nu$,
could be written as another, dual Chern-Simons action with coefficient
$1/k$ for particles of charge $1/k$ at filling fraction $\nu_d$. The caveats
necessary here are a) that their arguments on the irrelevance of other
terms generated in the dual action are compelling deep in the QH phases,
but do not take account of any anomalous dimensions that would be produced
near a $T=0$ critical point, and b) that the introduction of disorder should
be expected to change the precise functional connection between $\nu$ and
$\nu_d$. We also noted that for duality to be a symmetry in the sense of
Eqs. (\ref{eq:dualinear}), it is necessary that the dual actions lead to the
same $\rho^b(\nu)$.

The latter assumption was made by KLZ, formally based on a problematic
RPA treatment of the disorder which ignores any internal Chern-Simons
lines in the diagrams that contribute to $\rho^b$. This treatment was
subsequently called into question by
calculations on undisordered systems that suggested that the Chern-Simons
term is generically a marginal perturbation and gives rise of a line of
fixed points with continuously varying exponents \cite{against};
diagrammatically, they showed that the neglect of internal Chern-Simons
lines is not always justified.

Nevertheless, there is a second, qualitative argument in support of
KLZ's claim that appears more robust and has been the inspiration
for some recent, competing, model calculations for the defense
\cite{for}. If we accept the long wavelength form invariance
of the original and dual actions, then it follows that universal
quantities computed from them, internal gauge field lines and all,
can depend only upon the filling fraction (or the appropriate scaling field),
the charge of the bosons and the Chern-Simons coefficient. As both actions
describe the same transition,
it follows that they must yield the same correlation length/time exponents
despite the differing charge and statistics. That suggests, though it
does not dictate, that they yield the same scaling functions for the
resistivities; indeed, for particle-hole asymmetric disorder the converse
would seem to be a serious possibility (see later). At any rate it calls
into question the relevance of
calculations where the exponents {\em do} vary with the Chern-Simons
coupling. (These calculations \cite{against} perturb in the
Chern-Simons coefficient and hence are inconclusive on the large
coefficient shift $k \rightarrow 1/k$ at issue in the question of
duality. Recent model calculations \cite{for} have attempted
to show that models that do exhibit duality do not display any
statistics dependence of exponents at all.)

We should mention here also the work of L\"utken and Ross \cite{lr}
who were concerned with a description of QH systems on the basis
of actions in which the scale dependent resistivities appear as
parameters (along the lines of localization theory \cite{pruisken})
and postulated that a complexified duality or modular invariance
operates on them. Their {\em ansatz} is equivalent to that of KLZ for
our purposes.

\noindent
{\bf Particle-Hole Symmetry:} Within the framework of the fermionic
Chern-Simons theory, there are analogous issues to the ones discussed
above. However it is possible to gain some insight by studying the
problem of non-interacting electrons in a random potential which
is already non-trivial and at the very least may be a solution
(in the Hartree-Fock sense) of the {\em interacting} $\nu=1$ to insulator
transition, as suggested by some theoretical work \cite{lee-wang}.

In particular, one can check if microscopic asymmetries of the random
potential are irrelevant at the fixed point for the problem. In some
measure this was done by Huo, Hetzel and Bhatt \cite{hhb} in their
numerical studies which found that that critical conductivities were
insensitive to departures from microscopic particle-hole symmetry. The
contrast between their calculations of the density of states, a
microscopic quantity, and the long wavelength Hall conductivity
(see Figs. 2 and 3 in their paper) also support the conclusion that
departures from particle-hole symmetry are indeed
irrelevant. A second piece of evidence to this effect is the structure
of the network model of Chalker and Coddington \cite{network} which is
believed to be in the same universality class. The network model
clearly exhibits a symmetry between the QH and insulating phases; in the
absence of boundary effects inessential to the bulk physics, they are chirally
reversed translates of each other.

\noindent
{\bf Constancy of $\rho_H$:}
The possibility of a quantized Hall resistivity into the insulator
was implicit in the work of Jain and Kivelson \cite{jk} and KLZ who
first suggested the idea of a Hall insulator where $\rho_H$ would be
finite. This was later studied in detail by Dykhne and Ruzin
\cite{Ruzin1} and Ruzin and Feng  \cite{Ruzin2}. In this work
it is phrased in terms of a ``semi--circle law'' relating
$\sigma_{L}(\nu)$ and $\sigma_{H}(\nu)$. In the special case of a
transition from a $1/k$ QH state to an insulator, this law states
\beq
\sigma_{L}^2+\sigma_{H}^2=\sigma_{H}/k
\label{sclaw}
\eeq
and is equivalent to the constancy of $\rho_H$, while in other cases
it reflects the constancy of $\rho_H$ for the ``upper'' fluid defined
in the next section.
Eq. (\ref{sclaw}) was proved both for a classical two fluid model and
within the network model, assuming linear response.
Recently, it has been shown in a classicized
(i.e. non-interfering) version of the network model that the
constancy can persist beyond the linear response regime \cite{es-aa}.
As argued in previous chapters, in terms of composite bosons
this semi--circle law corresponds to a vanishing Hall resistance. It is
enlightening to point out, that a proof of $\rho_{H}^b=0$ is in fact
included in Ref. \cite{Ruzin2}, though in a different language. In
their formulation, they introduced  {\em local} current densities,
$\bf j_1$ and $\bf j_2$, for the two phases in the transition region and showed
that their average values are perpendicular to each other - a property
that is proved necessary and sufficient for the semi-circle law to be
obeyed. Translating this to the
bosonic Chern-Simons representation, liquids 1 and 2 correspond to mobile
bosons and vortices, hence $\bf j_1$ and $\bf j_2$ are the charge and vortex
current
densities. The latter has the significance of an electric field in the
perpendicular direction; Ruzin and Feng's statement therefore implies that
the current
and voltage in the bosonic description are {\it parallel} -- i.e., the Hall
coefficient vanishes!

Finally, we should note that for the similar problem of the field
tuned superconductor/insulator transition it has been argued by
Fisher \cite{mpaf-hall}, that an asymptotic particle-hole symmetry at
the critical point might lead to a vanishing Hall coefficient as suggested
by some data \cite{hebard}.
This does suggest the possibility that, in the QH system, both duality
and the vanishing Hall response might ultimately be consequences of the same
underlying principle.

\section{Duality Near Other Quantum Hall  Phase Transitions}

The hierarchical principle was invoked in Ref. \cite{klz} to
argue that {\em all} QH transitions are, in a precise sense,
transitions from principal QH states to insulators. Qualitatively,
they all consist of a ``lower'' (parent) fluid that is inert across
the transition and an ``upper'' (quasiparticle) fluid that undergoes
a transition to an insulator. For example the $\nu=2 \rightarrow 1$
transition is the $\nu=1 \rightarrow 0$ of the spin down lowest
Landau level while the spin up Landau level remains inert.

It follows that the symmetries of the transport observed in
\cite{expts} should be present at all transitions if one can
identify the transport coefficients of the upper fluid. This
is straightforward for the conductivities. As the fluids conduct
in parallel,
\beqarr
\sigma_{L}(E,\nu) &=& \sigma^u_{L}(E,\nu) \nonumber \\
\sigma_{H}(E,\nu) &=& \sigma^u_{H}(E,\nu) + \sigma^l_{H} \ ,
\label{EQ:lcssig}
\eeqarr
where the lower fluid contributes only a fixed, quantized Hall conductivity
independent of filling fraction and electric field \cite{fn-nonlinhall}.
As typical Hall bar measurements yield resistivities and, more importantly,
the symmetry itself would be manifest in the resistivities of the
upper fluid we need a prescription to go between them. In linear
response this is simple: we convert the measured $\rho_L$ and $\rho_H$ to
$\sigma_L$ and $\sigma_H$ by matrix inversion, obtain $\sigma^u_{L}$
and $\sigma^u_{H}$ from Eq. (\ref{EQ:lcssig}) and invert {\em them} to get
$\rho^u_L$ and $\rho^u_H$.

Beyond linear response one needs explicit expressions for the current
carried by the upper fluid, which is {\em not} parallel to the net current,
and for the electric fields resolved along and perpendicular to it.
Consider a Hall bar geometry where a current density $j$ flows along
the bar. Voltage measurements yield the longitudinal and Hall electric
fields $E_L, E_H$ and hence the (in general) non-linear resistivities
$\rho_L(j)$, $\rho_H(j)$. Let us denote the Hall angle between the
current and the total electric field of magnitude $E=\sqrt{E_L^2 +
E_H^2}$ by $\theta$, so that $E_L=E\cos\theta $, $E_H=E\sin\theta$.
Denoting the current density and Hall angle for the upper fluid
by $j^u$ and $\theta^u$, Eq.~(\ref{EQ:lcssig}) can be recast as
\beqarr
j\cos\theta&=&j^u\cos\theta^u \nonumber \\
-j\sin\theta&=&-j^u\sin\theta^u + \sigma^l E \ .
\label{EQ:jjo}
\eeqarr
The two equations in Eq.~(\ref{EQ:jjo}) suffice to determine $j^u$ and
$\theta^u$, and hence $E^u_L$ and $E^u_H$, in terms of the applied $j$ and
measured $\rho_L$ and $\rho_H$. After some algebra we obtain,
\beqarr
j^u &=& j \{ \frac{ \rho_L^2 +
[\rho_H + \sigma^u_H (\rho_L^2 + \rho_H^2)]^2}
{ \rho_L^2 + \rho_H^2 } \}^{1/2} \nonumber \\
E^u_L &=& \frac{j^2}{j^u} \rho_L \nonumber \\
E^u_H &=& \frac{j^2}{j^u} [\rho_H + \sigma^u_H (\rho_L^2 + \rho_H^2)]^2]
\ .
\eeqarr
Plots of $E^u_L$ and $E^u_H$ as functions of $j^u$ should then be expected
to resemble the $I$-$V_{L}$,  $I$-$V_{H}$ characteristics reported in
Ref. \cite{expts}.

\section{Summary and Open Questions}

In this paper we have followed a single line of argument in interpreting
the experimental results, i.e. we have assumed that the physics in the
transition region is governed by a zero temperature quantum critical
point. A great virtue of such an interpretation is that measured
quantities become properties of a scaling limit where it is possible
for symmetries, not manifest microscopically, to emerge because
the operators that break them are irrelevant at the underlying fixed
point. In other words, critical points can provide a robust rationale
for long wavelength symmetries. Nevertheless, we should remark that
this is not the only possibility. While the asymptotic low temperature
region (if it can be accessed on realistic time scales) for large samples
should be governed by critical physics one has to leave open the possibility
that the accessible temperature range might involve more complicated
finite temperature effects and lead to some of the same physics for more
classical reasons as in \cite{Ruzin1,es-aa}. For the samples studied to
date that exhibit the reflection symmetry, the temperature range that
shows evidence for scaling is too small for us to rule out such a
possibility.

With this caveat, we have argued that the reflection symmetry can
be naturally interpreted as duality/particle-hole symmetry combined
with a vanishing/constant Hall response in the composite boson/fermion
descriptions of the QH to insulator transitions. We expect that this
is a feature of all continuous QH transitions and have indicated
how to search for it at other transitions. We have also reviewed the
theoretical evidence in support of these inferences and find that
though not dispositive, it is certainly encouraging; at least for
non-interacting electrons, it is easy to visualize calculations
that can test them further \cite{lopez}. Overall, we find that the
experimental data through the interpretation of duality offer the
strongest evidence yet that KLZ's general framework of a universal
bosonic transition underlying the QH transitions is correct.

The fit between our analysis and the data, though compelling, is not
perfect. In particular, the derivation of the relationship between
$\nu_d$ and $\nu$ should be strictly valid only for a system that
exhibits duality down to the microscopic length scale,
e.g. for non-interacting electrons this is the case for microscopically
particle-hole symmetric disorder in the lowest Landau level.
For systems where duality is recovered only at long wavelengths,
the relationship should be more complicated. Indeed, in the strict
scaling limit, $T$ and $\nu-\nu_c \rightarrow 0$ with $x=(\nu-\nu_c)/
T^{1/\nu z}$ fixed, it matters very little what we pick. Nevertheless,
this does {\em not} mean that duality is without consequence. For
example, it would still imply that, as a function of the scaling variable
$x$, the current and voltage trade places at dual values and would have
the consequence that $\rho_{L}^c=1$. We feel that the correct perspective
on our analysis is that we have approximately identified a non-linear
scaling field ($\nu'-\nu'_c$ in the fermionic description) that allows the
symmetry
to be identified over a wider range of $\nu$ at accessible values of $T$.
It would however be very useful to get some quantitative understanding of
why $\nu^\prime-\nu^\prime_c$ continues to be such a good scaling field
even when $\nu'_c$ itself is shifted from its symmetric value of $1/2$
by as much as 20\%. This problem becomes more serious if the transition
out of the QH state is studied as a function of disorder at fixed filling.
While the formulation in terms of the scaling variable remains valid it
is not obvious how one might intepret any data, away from the scaling limit,
that might become available for this transition. A second problem, which is
difficult to pin down experimentally given the difficulty of accurately
measuring
$\rho_{H}$ deep into the insulator, is that the reflection symmetry for
the longitudinal response appears to hold over a larger range of fillings
than those over which the Hall response is constant. We do not have
a good understanding of this difference.

A different issue is the nature of the non-linear response. We have
suggested that the non-linear response is a consequence of intrinsic,
critical physics. It is not difficult to imagine getting the same
result from a heating scenario in which the electron gas equilibrates
at a different temperature from the lattice. What one needs for this
purpose is the reflection symmetry of the linear response plus an effective
electronic temperature that depends upon the dissipation in the bulk
alone, i.e. on the product $j E_L$ which is the same between dual points
on the longitudinal characteristics. The reason we are suspicious of
this mechanism is that it ignores the dissipation at the contacts,
where the Hall voltage is dropped and which is very asymmetric between
dual points. This is clearly an important issue and we expect to make
progress on it by further analysis of the data in the near future.
We should note that even if a heating scenario is correct, the symmetries
of the linear response would still require explanation.

Finally, we would like to draw attention to the possibility of
analogous symmetries near other interesting transitions. The most
closely related one is the field-tuned transition in two-dimensional
disordered superconducting films \cite{SITrev}. Here the theoretical
expectation \cite{mpaf-dual} is that it should {\em not} be
characterized by duality as the bosons and vortices interact with
different potentials. Unfortunately, the data here isn't completely
consistent, particularly on the question of universal values for
the critical resistivities; further studies with an emphasis on
the current-voltage characteristics might therefore be quite
useful. There is also the observation of a reflection symmetry
at the puzzling zero field transition observed by Kravchenko {\em
et al} \cite{kravchenko} which does not have any natural interpretation
in our framework. We expect these cross-comparisons to be very
instructive in evaluating the correctness of the analysis outlined
in this paper.


\acknowledgements
We thank A. Auerbach, Y. Avron, M. P. A. Fisher, E. Fradkin, D. E. Freed,
S. M. Girvin, S. A. Kivelson, M. Stone and D. C. Tsui for instructive
discussions. E.S. is grateful for the hospitality and support of the
Physics Department and ITP at the Technion, Israel, where part of this
work was carried out. This work has been supported by the Beckman Foundation,
the Aspen Center for Physics and the Israeli Academy of Sciences (ES), the
A. P. Sloan Foundation and NSF grant \# DMR-9632690 (SLS) and the NSF (DS).


\begin{references}
\bibitem{freeE}
Here $F$ is a functional of a dynamic field, hence the suggestion of a
free energy is not appropriate. However, as it plays the role of a dynamical
analogue of the equilibrium energy functional, we hereon use this term
and drop the quotation marks.
\bibitem{fn-edgedisp} If the ideal regions have non-linear dispersions then
the Landauer conductances are not properties of the disordered regions alone
and the scheme is no longer sensible for our purposes.
\bibitem{fn-nonlinhall} There are certainly deviations from the ideal
response even for an ideal QH state, but they are far smaller than the
conductivities of the upper fluid and can be neglected in comparison.
\end{references}
\end{document}